# Direct observation of the thickness distribution of ultra thin AlO$_x$ barrier in Al/AlO$_x$/Al Josephson junctions

L. J. Zeng[1], S. Nik[1], T. Greibe[2], C. M. Wilson[2], P. Delsing[2], and E. Olsson[1,*]

[1]*Department of Applied Physics,
Chalmers University of Technology, Gothenburg, 41296, Sweden*
[2]*Department of Microtechnology and Nanoscience (MC2),
Chalmers University of Technology, Gothenburg, 41296, Sweden*

We show that less than 10% of the barrier area dominates the electron tunneling in state-of-art Al/AlO$_x$/Al Josephson junctions. They have been studied by transmission electron microscopy, specifically using atomic resolution annular dark field (ADF) scanning transmission electron microscopy (STEM) imaging. The direct observation of the local barrier thickness shows a Gaussian distribution of the barrier thickness variation along the junction, from ~1 nm to ~2 nm in the three junctions we studied. We have investigated how the thickness distribution varies with oxygen pressure ($p_o$) and oxidation time ($t_o$) and we find, in agreement with resistance measurements on similar junctions, that an increased $t_o$ gives a thicker barrier than an increased $p_o$.

PACS numbers: 74.50.+r, 73.40.-c, 74.81.-g, 85.25.-j

## INTRODUCTION

Aluminum oxide (AlO$_x$) layers of nanometer thickness are widely used as tunnel barriers in tunnel junctions that have a wide range of applications in devices such as radiation detectors, single-electron transistors, and superconducting qubits [1-6]. Magnetic tunnel junctions also often include ultrathin AlO$_x$ layers as the tunnel barrier between two magnetic layers [7]. The quality and structural characteristics of these oxide barriers may directly determine the tunneling characteristics of the charge carriers across the barrier and consequently the performance of the junctions. The most relevant physical parameter for characterizing the structure of the oxide barrier is the barrier thickness, since the tunneling current decreases exponentially with increasing barrier thickness. The thinnest region in the barrier will give rise to the highest tunnel current and become the preferential tunneling channel for the charge carriers. Consequently, the distribution of the barrier thickness will result in inhomogeneity of the tunnel current across the barrier, and the conductance per unit area becomes much smaller than if the tunnel barrier had been equally thick everywhere. In several experiments there is indirect evidence that only a small fraction of the junction area is active [8-10]. A tunnel barrier with varying thickness in a superconducting tunnel junction changes the ratio between multiparticle currents, based on Andreev processes, and the single-particle current. This leads to excess current in the subgap region where single-particle tunneling is suppressed. Excess subgap current has been found in various superconducting tunnel junctions and may become a limitation for the future applications of electron pumps [11, 12]. The origin of excess subgap current has been suggested to be the variation of the local junction transparencies resulting from a distribution of barrier thickness, or the existence of 'pinholes' in the junctions [13-16]. However, more recent studies based on low-temperature measurements of current-voltage characteristics of superconducting tunnel junctions have ruled out 'pinholes' as an explanation for the excess current but confirmed the role of the uneven distribution of the barrier thickness [10].

If the local transparency fluctuates in time, this may also lead to resistance noise in the tunnel junction and to critical-current noise in Josephson junctions. The latter may limit the coherence times in superconducting qubits [17,18].

Transport measurements and theoretical modeling have provided strong indications of the influence of barrier thickness on the performance of the tunnel junctions, and it is therefore of great importance to directly measure the thickness distribution of the oxide barriers in the junctions in order to evaluate the effects of detailed, local structure on the transport properties of the junction.

Here we report on a cross-section transmission electron microscopy (TEM) study on the microstructure of the AlO$_x$ barriers in Al/AlO$_x$/Al Josephson junctions with the main focus on the direct measurement of the barrier thickness distribution. While techniques such as conductive atomic force microscopy (CAFM), scanning tunneling microscopy (STM) and X-ray photoelectron spectroscopy (XPS) [19-22] are usually used to measure the barrier thickness, the information they provide is usually indirect and/or has relatively low spatial resolution. CAFM and STM are used to measure the tunnel current locally (~ nm$^2$ scale) and then derive the tunnel barrier parameters (barrier thickness, barrier height etc.) by fitting the current-voltage curve with some physical model. For XPS measurements, the results are usually averaged over a sample surface area with the size in ~ µm$^2$ scale. In this study, with atomic resolution scanning TEM (STEM) imaging, we directly show a local thickness variation of the barrier thickness along the junction and also that partial oxygen pressure ($p_o$) and oxidation time ($t_o$) during fabrication affect the barrier thickness distribution. The discussion also extends to the tunnel current distribution as a result of the variation in the barrier thickness. The results provide important insights for the understanding of the transport properties of superconducting devices such as excess subgap conductance in tunnel junctions and decoherence in qubits. It could also shed light on the pathways whereby further improved tunnel barriers can be produced with respect to the homogeneity in barrier thickness.

## EXPERIMENTS

The samples used in this study were grown on $SiO_2$/Si substrates in high vacuum by thermal evaporation. A bottom Al film of nominal thickness of 15 nm was deposited with a deposition rate ranging from 9 to 12 Å/s. The Al film was thereafter exposed to high-purity (99.99%) $O_2$ with fixed pressure and time. Subsequently the top Al layer with a nominal thickness of 60 nm was deposited with the same deposition rate as that for the bottom Al film. The tunneling characteristics of the junctions used in this study are representative of large numbers of the junctions analyzed in Ref. [10]. Three sets of unpatterned samples were used in this study. They were made with the same film deposition parameters but with different barrier oxidation parameters, namely, with $p_o$&$t_o$ being 0.1 mbar&3 min (sample 1), 0.1 mbar&30 min (sample 2), and 1 mbar&3 min (sample 3) respectively, as shown in Table I.

Cross-section TEM specimens were prepared by grinding and polishing the specimen down to ~ 20 μm, followed by Ar ion milling. The specimens were kept at about -80°C during milling to minimize the damage from the ion beam. An FEI Titan 80-300 TEM/Scanning TEM (STEM) was used for high resolution imaging. Annular dark field (ADF) STEM images were acquired with a 19.7 mrad beam convergence angle and 54-270 mrad detector collection angle. The spatial resolution of the microscope in ADF STEM mode is determined to be ~1Å.

|  | $p_o$ (mbar) | $t_o$ (minute) | $\langle l \rangle$ (nm) | $\sigma_l$ (nm) |
|---|---|---|---|---|
| Sample 1 | 0.1 | 3 | 1.66 | 0.351 |
| Sample 2 | 0.1 | 30 | 1.88 | 0.326 |
| Sample 3 | 1 | 3 | 1.73 | 0.372 |

TABLE I. Oxidation parameters (partial oxygen pressure ($p_o$) and oxidation time ($t_o$)) and barrier thickness measurement results (average barrier thickness ($\langle l \rangle$) and standard deviation ($\sigma_l$)) for the three samples analyzed in this study.

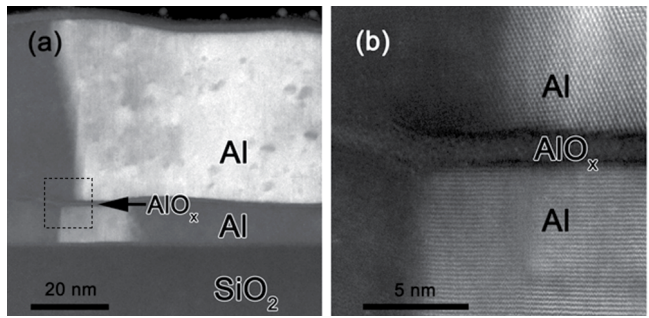

FIG. 1 ADF STEM images of an Al/AlO$_x$/Al tunnel junction. (a) An overview of the cross-section of the junction showing the Al electrodes, AlO$_x$ barrier and SiO$_2$ substrate layer. (b) An ADF image acquired from the dashed area in (a).

Figs. 1(a) and (b) show typical cross-section STEM annular dark field (ADF) images of the junction (sample 1). Different layers (SiO$_2$ substrate, bottom Al electrode, AlO$_x$ barrier, and top Al electrode) in the junction are indicated in the images and the film growth direction is from bottom to top. In Fig. 1(a) we can clearly see the polycrystalline and columnar nature of both top and bottom Al electrodes. Because of the channeling effect, there is a higher intensity in the ADF image when the electron beam travels along the zone axis of the Al grains compared to when the incident beam direction is different from the zone axis. Thus, the varying intensity in the Al layers in the images is due to Al grains with different orientations. The grain size of the Al in the junction can therefore be measured from the images. The Al grain size measured along the direction perpendicular to the film growth direction varies from 20 nm to 50 nm for the bottom Al electrode, and from 80 nm to 200 nm for the top Al electrode. There is a dependence of the grain size of the top layer Al on the oxidation time. A more detailed grain size distribution analysis will be reported elsewhere [23]. Along the film growth direction, each of the Al grains in both the bottom and top Al layers extend throughout the film thickness as a consequence of the columnar growth. The AlO$_x$ barrier layer follows the top surface of the bottom Al layer, where the surface roughness differs from area to area. No voids or pinholes were observed in any of the samples we have investigated. Fig. 1(b) shows a high magnification STEM image taken from the area indicated by the dashed frame in Fig. 1(a). The atomic columns and lattice fringes of the Al grains and the Al/AlO$_x$ interface structure can be clearly seen in the image.

Because of the incoherent nature of STEM ADF imaging, Fresnel fringes and other interference phenomena will not appear in the ADF image and degrade the accuracy of the barrier thickness measurement [24]. STEM ADF images were thus used for barrier thickness measurements.

Fig. 2(a) shows a high resolution ADF STEM image of the junction from sample 3. The atomic columns and lattice fringes from Al grains on both sides of the amorphous barrier are clearly visible in Fig. 2(a). In such an ADF STEM image, brighter contrast normally corresponds to atomic columns and

## RESULTS AND DISCUSSION

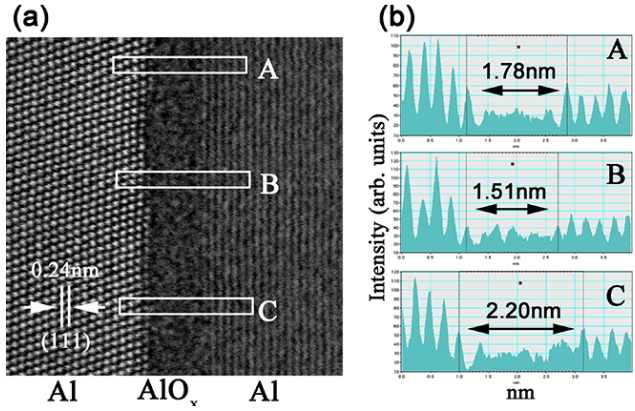

FIG. 2. (a) A high resolution ADF STEM image of the junction from sample 3. The Al (111) atomic plane and the plane distance are indicated. Windows A, B and C with size of ~ 4*0.5 nm² in the image show the areas used to measure the intensity profile of the image. Intensity profiles are measured across the barrier and integrated along the Al/AlO$_x$ interface. (b) Image intensity profiles acquired from area A, B and C that are marked in (a). Barrier thicknesses measured from the intensity profiles are also shown.

lattice fringes in the material because in ADF STEM mode we are collecting electrons that are scattered by the atomic columns to a certain angle. The space between the atomic columns contributes much less to the scattering of incoming electrons so in the image it appears dark. From Fig. 2(a), we can also examine the detailed interface structure at the Al/AlO$_x$ interface, e.g., there are atomic steps of Al at the interface, which can contribute to the variation of barrier thickness at the atomic scale. Intensity profiles at 3 positions (A, B, and C) in the image were measured to demonstrate how the barrier thickness measurement is done and how the barrier thickness varies. At each position, the intensity profile across the barrier was measured and integrated along the direction parallel to the Al/AlO$_x$ interface. Each measurement was done using a window with the size of ~ 4*0.5 nm² as shown in Fig. 2(a). The corresponding intensity profiles are shown in Fig. 2(b). The intensity profiles from Al show sharp peaks corresponding to the positions of Al atomic planes. The distance between two neighboring peaks is the atomic plane distance, which is indexed and marked in Fig. 2(a). In contrast, the intensity profile from the AlO$_x$ barrier has a relatively random intensity distribution. Thus, we can determine the interface between the crystalline Al layers and the amorphous AlO$_x$ barrier from the intensity profiles and consequently measure the barrier thicknesses. The pixel size of Fig. 2(a) is 0.019 nm. We estimate that the uncertainty in the measurement is around 0.04 nm. As shown in Fig. 2(b), the barrier thickness was measured to be 1.78 (±0.04), 1.51 (±0.04), and 2.20 (±0.04) nm at position A, B, and C respectively.

We measured the thickness of the barrier at over 300 different positions for each sample in order to measure the barrier thickness distribution. For each sample, we used around 50 STEM images from approximately 20 different grains for the thickness measurement. In each high resolution STEM image we measured the barrier thicknesses at positions at a distance around 2 nm away from each other. The resulting barrier thickness distributions for the three samples are shown in Fig. 3. As shown in Fig. 3(a), for the barrier oxidized with $p_o$ = 0.1 mbar and $t_o$ = 3 minutes (Sample 1) the barrier thickness varies from 1.2 nm to 2.2 nm. More than sixty percent of the barrier thickness is in the range from 1.5 to 1.8 nm, and the mean barrier thickness is 1.66 nm. A comparison between the barriers in Sample 1 and Sample 3 shows that the mean barrier thickness increases from 1.66 nm to 1.73 nm when $p_o$ was increased by a factor of 10. There is a relatively larger change of mean barrier thickness when the $t_o$ is increased from 3 minutes for Sample 1 to 30 minutes for Sample 2, even though the change is also rather small (from 1.66 nm to 1.88 nm). We can compare this with statistics from junction resistances depending on oxidation parameters, which we present in Fig. 4. Defining an effective oxidation dose $D = t_o^{0.65} * p_o^{0.43}$, we find that for a large number of samples, the conductance per unit area decreases linearly with D. The two exponents have been optimized to minimize the spread of the data. The measured average thicknesses presented in Table I agree with this and confirms that the oxidation time is more important than oxidation pressure, that is, a tenfold increase in $t_o$ gives a thicker barrier than a tenfold increase of $p_o$.

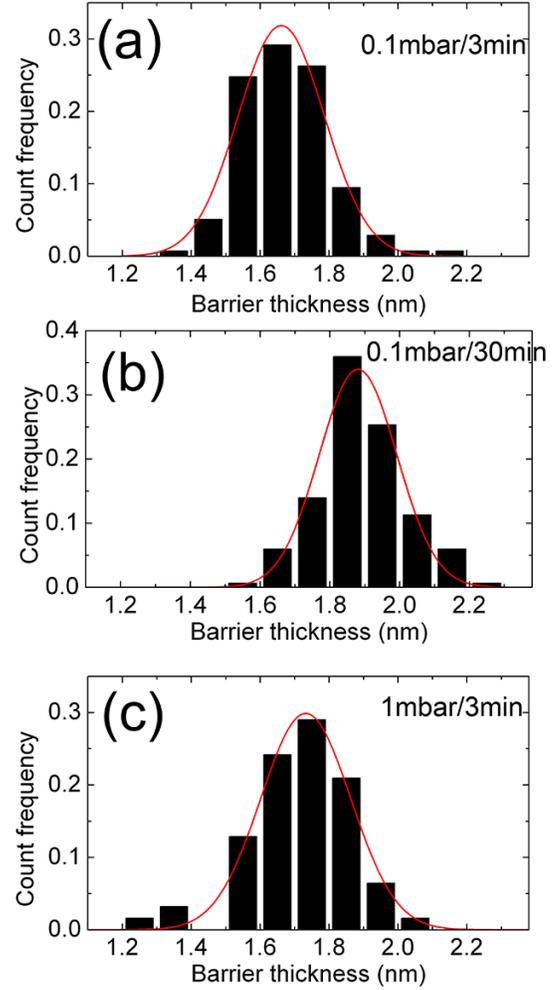

FIG. 3. Barrier thickness distributions measured from ADF STEM images of three AlO$_x$ barriers made at different oxidation conditions: (a) $p_o/t_o$~0.1 mbar/3 min (sample 1), (b) $p_o/t_o$~0.1 mbar/30 min (sample 2), (c) $p_o/t_o$~1 mbar/3 min (sample 3). Experimental thickness distributions are also fitted with Gaussian functions, shown as the red curves superimposed on the experimental data.

The variation of the mean barrier thickness shown here is, however, different from the result of previous XPS measurements [22], in which the thickness of $AlO_x$ thin film increased more rapidly with $t_o$. Moreover, the inhomogeneity of the barrier thickness of the junction, ranging from ~1 nm to ~2 nm, was observed in all the samples we studied, as shown in Fig. 3. The thickness inhomogeneity has also been derived from CAFM and STM measurements [19-21]. In contrast, we have directly measured the barrier thickness with atomic resolution in TEM. The tunneling probability of charge carriers across the barrier is an exponential function of the barrier thickness. It has been shown that a 0.2 nm decrease of barrier thickness could result in one order of magnitude increase in tunneling current [21]. Thus, the thinnest region in the barrier may act as an active region or 'hot spot' for tunneling. The distribution of the barrier thickness is also fitted with a Gaussian distribution for each sample. The fitting results are shown by the curves superimposed on the histograms in Fig. 3. The fitting gives the average thickness $\langle l \rangle$ for each sample and the standard deviation $\sigma_l \ll \langle l \rangle$ of the barrier thickness. $\langle l \rangle$ is 1.66 nm, 1.88 nm and 1.73 nm and $\sigma_l$ is 0.351 nm, 0.326 nm and 0.372 nm for Sample 1, 2 and 3 respectively. A recent study of $Cu/AlO_x/Al$ junctions also shows a similar barrier thickness distribution [25].

We note that though no pinhole-like microstructural defects were observed, the strong fluctuation of barrier thickness in the junction is likely to give rise to a variation in transmittance in the barrier. It, thus, could contribute to the excess subgap current in $Al/AlO_x/Al$ tunnel junctions via a multiple-particle tunneling process in the thin area [13] e.g., Andreev transport [10].

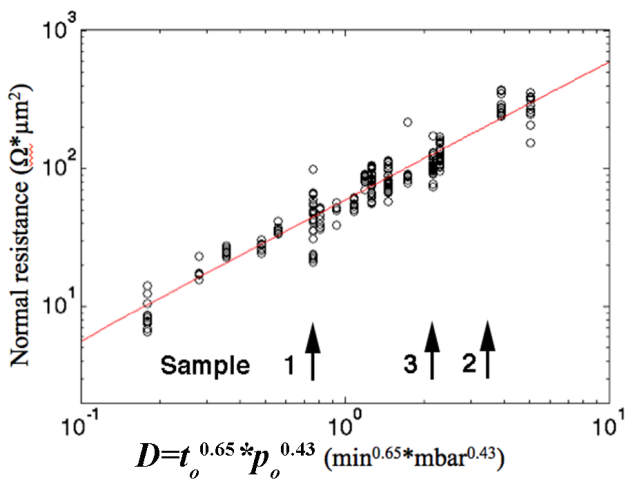

FIG. 4. Dependence of normal resistance on oxidation parameters. The X-axis represents an effective oxygen dose defined as $D = t_o^{0.65} * p_o^{0.43}$. Black circles show the experimental measurement results of normal resistance ($R_n$ times junction area) of $Al/AlO_x/Al$ junctions prepared at different oxidation conditions. The red line shows the linear fit to the experimental data.

Based on the barrier thickness distribution measurements, we further calculated the distribution of the tunnel current as a function of the barrier thickness. The tunnel current through a rectangular potential barrier of height $\phi$ and thickness $l$ can be expressed as $J = V * A * \exp[-l/\lambda]$, where $V$ is the voltage applied across the junction and can be set as a constant in this case, $\lambda = \hbar / 2\sqrt{2m\phi}$ is the attenuation length in the barrier and $A = \left[3*(2m\phi)^{1/2}/2l\right]*(e/h)^2$ [26]. $\phi$ is chosen to be ~ 2 V, which is a typical value for $AlO_x$ barriers [20, 21, 27].

Thus the distribution function for the tunnel current in a tunnel junction is $T(l) = J(l) * g(l)$, where $g(l)$ is the thickness distribution function obtained by fitting the experimental data to a Gaussian distribution as shown in Fig. 3. In order to investigate the proportion of the most active area in terms of electron tunneling in the junctions, we define $P(l_x) = \int_{-\infty}^{l_x} T(l) dl \Big/ \int_{-\infty}^{+\infty} T(l) dl$. We also define $G(l_x) = \int_{-\infty}^{l_x} g(l) dl \Big/ \int_{-\infty}^{+\infty} g(l) dl$ to calculate the proportion of the barrier area with the thickness smaller than $l_x$ with respect to the whole barrier area in the junction.

$P(l_x)$ as a function of $G(l_x)$ for the three samples we investigated are calculated. We find that in sample 1, $G(l_x)$ is around 0.73 when $P(l_x) = 0.9$. It means 90% of the tunnel current in the whole junction is from about 7.3% of the junction area. Similarly, in sample 2, 90% of the total tunnel current is from 9.2% of the junction area. In sample 3, 90% of the total tunnel current comes from around 6.9% of the junction area. Under this criterion, based on our measurements and calculations shown in Fig. 3, less than 10% of the barrier area is active in each barrier. This finding is consistent with the assumptions made for theoretical modeling of Superconductor-Insulator-Superconductor and Normal metal-Insulator-Superconductor junctions [8, 10].

The total junction conductance is given by the full integral, which for a Gaussian distribution becomes: $G(d,\sigma) = G_0 e^{-d/d_0} e^{\sigma^2/(2d_0^2)}$. We see that for a homogeneous distribution (s=0) we recover the original exponential dependence, but for finite s we get an increased conductance.

There are several factors that may contribute to the variation of the thickness of the $AlO_x$ barrier oxidized directly on the Al bottom layer: (i) Normally, the oxide barrier closely follows the morphology of the top surface of the bottom Al layer. However, the barrier tends to be thicker at some Al grain boundaries because of the grain boundary grooving [28]; (ii) The bottom Al layer is polycrystalline, so the Al grains have different crystallographic orientations along the film growth direction. Even on one single grain, the curvature of the top surface of the Al grain indicates that there might be local variation in the crystallographic orientation of Al. Al grains with different crystallographic orientations have different surface energy and also different atomic plane distance along the top surface and perpendicular to the surface. When oxidized, it will thus give rise to a variation in oxide thickness [22, 29]; (iii) Atomic steps of Al at the $Al/AlO_x$ interfaces result in local change of the barrier thickness.

# CONCLUSION

In summary, the barrier thickness distribution of the AlO$_x$ barrier in state-of-art Al/AlO$_x$/Al tunnel junctions has been investigated by transmission electron microscopy at atomic resolution. There is a small change of the mean barrier thickness for the barriers made with $p_o/t_o$ of 0.1 mbar/3 min, 0.1 mbar/30 min and 1 mbar/3 min. STEM imaging shows that the barrier thickness has a distribution, ranging from ~1 nm to ~2 nm, in every aluminum oxide barrier we have studied. The average barrier thickness confirms previous experimental results that time is a more important parameter than pressure during the oxidation. According to the barrier thickness distribution measurements and tunnel probability calculation, less than 10% of the total barrier area is active in the tunneling process in all three junctions studied. The results show the strong inhomogeneity of tunneling current resulting from the barrier thickness distribution.

# ACKNOWLEDGEMENTS

We thank the Swedish Foundation for Strategic Research, the Swedish Research Council, the Knut and Alice Wallenberg Foundation and Chalmers Nanotechnology Center for financial support.

*eva.olsson@chalmers.se
[1] J. Zmuidzinas, P.L. Richards, Proceedings of the IEEE, 92, 1597 (2004).
[2] K.K Likharev, Proceedings of the IEEE, 87, 606 (1999)
[3] J. Clarke and F.K. Wilhelm, Nature 453, 1031-1042 (2008)
[4] N. E. Booth and D. J. Goldie, Supercond. Sci. Technol. 9, 493 (1996).
[5] T. J. Harvey, D. A. Rodrigues, and A. D. Armour, Phys. Rev. B 78, 024513 (2008).
[6] F. Paauw, A. Fedorov, C. Jarmans, and J. Mooij, Phys. Rev. Lett. 102, 090501 (2009).
[7] J. S. Moodera, L. R. Kinder, T.M. Wong, and R. Meservey, Phys. Rev. Lett. 74, 3273 (1995).
[8] H. Pothier, S. Guéron, D. Esteve, and M. H. Devoret, Phys. Rev. Lett. 73, 2488 (1994).
[9] V. F. Maisi, O.-P. Saira, Yu. A. Pashkin, J. S. Tsai, D. V. Averin, and J. P. Pekola, Phys. Rev. Lett. 106, 217003 (2011).
[10] T. Greibe, M. P.V. Stenberg, C. M. Wilson, T. Bauch, Vitaly S. Shumeiko, and P. Delsing, Phys. Rev. Lett. 106, 097001 (2011).
[11] J. P. Pekola, J. J. Vartiainen, M. Möttönen, O.-P. Saira, M. Meschke and D. V. Averin, Nature Phys. 4, 120 (2008).
[12] D. V. Averin and J. P. Pekola, Phys. Rev. Lett. 101, 066801 (2008).
[13] J. R. Schrieffer and J.W. Wilkins, Phys. Rev. Lett. 10, 17 (1963).
[14] A.W. Kleinsasser et al., IEEE Trans. Appl. Supercond. 5, 2735 (1995).
[15] E. V. Bezuglyi, A. S. Vasenko, E. N. Bratus, V. S. Shumeiko, and G. Wendin, Phys. Rev. B 73, 220506(R) (2006).
[16] Y. Naveh, Vijay Patel, D.V. Averin, K. K. Likharev, and J. E. Lukens, Phys. Rev. Lett. 85, 5404 (2000).
[17] V. Zaretskey, B. Suri, S. Novikov, F. C. Wellstood, and B. S. Palmer, Phys. Rev. B 87, 174522 (2013).
[18] R. McDermott, IEEE Trans. Appl. Supercond. 19, 2 (2009).
[19] W. H. Rippard, A. C. Perrella, F. J. Albert, and R. A. Buhrman, Phys. Rev. Lett. 88, 046805 (2002).
[20] V. Da Costa, C. Tiusan, T. Dimopoulos, and K. Ounadjela, Phys. Rev. Lett. 85, 876 (2000).
[21] E. Z. Luo, S. K. Wong, A. B. Pakhomov, J. B. Xu, I. H. Wilson, and C. Y. Wong, J. Appl. Phys. 90, 5202 (2001).
[22] L.P.H. Jeurgens, W. G. Sloof, F. D. Tichelaar, and E. J. Mittemeijer, J. Appl. Phys. 92, 1649 (2002).
[23] S. Nik, T. Greibe, H. Pettersson, L. J. Zeng, P. Krantz, P. Delsing and E. Olsson (submitted).
[24] P. D. Nellist, and S. J. Pennycook, Adv. Imag. Electron Phys. 113, 148-203 (2000).
[25] T. Aref, A. Averin, S. van Dijken, A. Ferring, M. Koberidze, V. F. Maisi, H. Nguyen, R. M. Nieminen, J. P. Pekola, and L. D. Yao (submitted).
[26] J. G. Simmons, J. Appl. Phys. 34, 1793 (1963).
[27] K. Gloos, P. J. Koppinen and J. P. Pekola, J. Phys. Condens. Matter 15, 1733 (2003).
[28] G. L. J. Bailey and H. C. Watkins, Proc. Phys. Soc. B 63, 350 (1950).
[29] L. P. H. Jeurgens, W. G. Sloof, F. D. Tichelaar and E. J. Mittemeijer, Thin Solid Films 418, 89 (2002).